\documentclass[a4paper,11pt]{article}

\usepackage[utf8x]{inputenc}
\usepackage[centertags]{amsmath}
\usepackage{amssymb,amsthm,epsfig,psfrag}
\usepackage{graphicx}
\usepackage{dsfont}
\usepackage{color}
\usepackage{pstricks-add}
\usepackage{fullpage}
\usepackage{etex}

\usepackage{bm}
\usepackage{epic}
\usepackage{hyperref}
\usepackage{young}
\usepackage{multirow}
\usepackage{fancybox}

\newcommand{\sfrac}[2]{{\textstyle\frac{#1}{#2}}}
\newcommand{\osca}{\mathbf{a}}
\newcommand{\oscb}{\mathbf{b}}
\newcommand{\oscab}{{\bar{\mathbf{a}}}}
\newcommand{\oscbb}{{\bar{\mathbf{b}}}}

\newcommand{\oscw}{\mathbf{w}}
\newcommand{\oscwb}{{\overline{\mathbf{w}}}}

\usepackage{fix-cm}

\usepackage[nosort]{cite}
\usepackage[bulletsep]{collref}
\usepackage{tensor}

\begin{document}

\thispagestyle{empty}

\begingroup\raggedleft\footnotesize\ttfamily
HU-EP-14/11\\ 
AEI-2014-007\\
HU-Mathematik-2014-04\\

\vspace{15mm}
\endgroup

\begin{center}
{\Large\bfseries A Shortcut to General Tree-level Scattering Amplitudes\\ in $\mathcal{N}=4$ SYM via Integrability\par}%
\vspace{15mm}

\begingroup\scshape\large 
Nils Kanning${}^{1,2}$, Tomasz \L ukowski${}^{3}$,
Matthias Staudacher${}^{1,2}$
\endgroup
\vspace{7mm}

\textit{
${}^1${
Institut f\"ur Mathematik, Institut f\"ur Physik und IRIS Adlershof,\\
Humboldt-Universit\"at zu Berlin,\\
Zum Gro\ss en Windkanal 6, 12489 Berlin, Germany}\\[0.2cm]
${}^2${
Max-Planck Institut f\"ur Gravitationsphysik, Albert-Einstein-Institut,\\
Am M\"uhlenberg 1, 14476 Potsdam, Germany
}\\[0.2cm]
${}^3${
Mathematical Institute, University of Oxford, Andrew Wiles Building,\\ Radcliffe Observatory Quarter, Oxford, OX2 6GG, United Kingdom}}\\[0.4cm]

{\tt kanning,staudacher$\bullet$mathematik.hu-berlin.de,\\
lukowski$\bullet$maths.ox.ac.uk}\\[0.2cm]

\vspace{8mm}

\textbf{Abstract}\vspace{5mm}\par
\begin{minipage}{14.7cm}

We combine recent applications of the two-dimensional quantum inverse scattering method to the scattering amplitude problem in four-dimensional $\mathcal{N}=4$ Super Yang-Mills theory. Integrability allows us to obtain a general, explicit method for the derivation of the Yangian invariants relevant for tree-level scattering amplitudes in the $\mathcal{N}=4$ model.

\end{minipage}\par
\end{center}
\newpage

\section{Introduction}
For now many years there has been considerable interest in the integrable properties of planar gauge theories. Of special importance as the primary playground for testing ideas of integrability is the planar $\mathcal{N}=4$ super Yang-Mills (SYM) model, the unique, maximally supersymmetric theory in four dimensions, for a comprehensive review see \cite{Beisert:2010jr}. Integrability has already proven to be a good tool for calculating many observables for that model, including the set of anomalous dimensions of composite operators as well as certain families of structure constants. The former is believed to be understood at any finite value of the coupling constant, the most advanced method for calculating them taking the form of a ``Quantum Spectral Curve'' \cite{Gromov:2013pga}. Recently, the attention of most workers in the field turned to the application of integrability methods to further quantities of interest, such as the expectation values of Wilson loops as well as scattering amplitudes. At strong coupling it already  turned out to be very useful, allowing to recast the leading part of scattering amplitudes in terms of a Y-system \cite{Alday:2010vh}. There have also been impressive advances towards the exact computation of Wilson loops and of amplitudes at any coupling \cite{Basso:2013vsa,Basso:2013aha,Basso:2014koa}. For example, the non-perturbative answer for expectation values of polygonal Wilson loops was reformulated as a sum over infinitely many particle contributions, which are, in principle, accessible via an asymptotic Bethe ansatz at any coupling. This provides a somehow orthogonal expansion compared to perturbation theory, giving rise to new predictions for the all-loop answers. There have also been important advances in understanding amplitudes using integrability at weak coupling. At tree-level the Yangian symmetry of amplitudes was proven in \cite{Drummond:2009fd}, which combined the invariance under explicit superconformal transformations of the model with its hidden counterpart -- a second, dual superconformal symmetry \cite{Drummond:2008vq}. 
This Yangian invariance, underlying a large class of rational two-dimensional integrable models, usually does not provide an immediate tool for calculations. However, it allows for a powerful approach termed the Quantum Inverse Scattering Method (QISM), where one constructs a family of operators in involution.

For the tree-level amplitudes, the first step towards the application
of the QISM was done in \cite{Ferro:2012xw}, where the crucial notion
of a spectral parameter was introduced to the scattering amplitude
problem in $\mathcal{N}=4$ SYM, see also \cite{Ferro:2013dga}. This
parameter found an interesting interpretation as a deformed, and thus
in general unphysical, particle helicity. With the use of on-shell
diagrams \cite{ArkaniHamed:2012nw}, providing a solution to the BCFW
recursion relation \cite{Britto:2005fq}, it allowed to deform
tree-level amplitudes in such a way that they satisfy (generalized)
Yang-Baxter equations. The latter are often interpreted as the
quintessence of quantum integrable models. A full classification of
Yangian invariants was also found in \cite{ArkaniHamed:2012nw}, based
on a deep relation between on-shell diagrams and permutations due to
Postnikov \cite{Postnikov:2006kva}. This relation to permutations was
further studied in \cite{Beisert:2014qba}, and will be an essential
ingredient in the construction proposed in this paper. The next steps
in putting scattering amplitudes into the QISM framework were
independently performed in \cite{Frassek:2013xza,Chicherin:2013ora},
where it was proposed to study certain auxiliary spin chain
monodromies built from local Lax operators. In this approach the
amplitudes are found as ``eigenstates'' of these monodromies. The
monodromies depend on an ``auxiliary'' spectral parameter, while the
spectral parameters of \cite{Ferro:2012xw,Ferro:2013dga} are encoded
as inhomogeneities of the Lax operators. The amplitudes do not depend
on the auxiliary spectral parameter, which is a key feature of the
QISM. In \cite{Frassek:2013xza} most details were given for a toy
version of scattering amplitudes, where the complexified
superconformal algebra $\mathfrak{gl}(4|4)$ was simplified to the
$\mathfrak{gl}(2)$ case. It was then shown how to obtain the Yangian
invariants from the monodromy matrix eigenproblem. Applying a modified
version of the Algebraic Bethe Ansatz, this led to a system of Bethe
equations for Yangian invariants.  A different, more direct, and very
powerful method, also based on the monodromy eigenproblem, was
proposed in \cite{Chicherin:2013ora}. However, neither in
\cite{Frassek:2013xza} nor in \cite{Chicherin:2013ora} a systematic
classification of Yangian invariants was provided. In this paper, we
would like to fill this gap, and detail an integrability-based
construction method for all Yangian invariants relevant to the
tree-level scattering amplitudes in $\mathcal{N}=4$ SYM. As a
byproduct, some interesting relations between the techniques in
\cite{Frassek:2013xza} and \cite{Chicherin:2013ora} as well as the
observations in \cite{Beisert:2014qba} will emerge. After completing this project we became aware of \cite{Broedel:2014pia}, which shares many conclusions with our work.

The paper is organized as follows. In section~\ref{Sec:Construction}
we begin by recapitulating some basic results of
\cite{Frassek:2013xza} and \cite{Chicherin:2013ora}, and we provide a
link between the two approaches. This link helps to understand how to
systematically generalize the powerful construction method of
\cite{Chicherin:2013ora} to general Yangian invariants. This
generalized construction is provided at the end of the section. As in
\cite{Frassek:2013xza}, we will, for pedagogical reasons, mostly
restrict the discussion in section~\ref{Sec:Construction} to
$\mathfrak{gl}(2)$ or certain closely related compact representations
of $\mathfrak{gl}(N|M)$. In section~\ref{Sec:Amplitudes} we explain
how the construction of the previous section generalizes, with small
changes, to the problem of deformed $\mathfrak{psl}(4|4)$ invariant
tree-level scattering amplitudes in $\mathcal{N}=4$ SYM.  In section
\ref{Sec:Examples} we illustrate how our method works for particular
examples with up to five external particles. We end with a summary and
outlook.

\section{Details of Construction}\label{Sec:Construction}

\subsection{Introductory Remarks}
The purpose of this paper is the systematic classification of Yangian
invariants relevant for the tree-level scattering amplitudes of
$\mathcal{N}=4$ SYM. Yangian invariance can be defined in a very
compact form as a system of eigenvalue problems for the elements of a
suitable monodromy matrix $M(u)$, cf.\ \cite{Frassek:2013xza},
\begin{equation}\label{invariance}
  M_{ab}(u)|\Psi \rangle=\delta_{ab}\,|\Psi \rangle\,.
\end{equation}
We are looking here for eigenvectors $|\Psi \rangle$ that are elements
of the space $V=V_1\otimes \ldots\otimes V_n$ with $V_i$ being a
representation space of a particular $\mathfrak{gl}(N|M)$
representation. The representations we are interested in have the
property that they can be built using a single family of harmonic
oscillators transforming in the fundamental representation of
$\mathfrak{gl}(N|M)$. In order to make our discussion more
transparent, we focus first on the $\mathfrak{gl}(2)$ algebra and
consider only compact representations. Later on we proceed to the
general problem with emphasis on the case $N|M=4|4$, relevant for the
$\mathcal{N}=4$ SYM amplitudes.

Following \cite{Frassek:2013xza} we distinguish two oscillator
realizations of the $\mathfrak{gl}(2)$ algebra,
\begin{align}\label{symm}
  J_{ab}&=+\oscab_a \osca_b&
  &\text{with}&
  [\osca_a,\oscab_b]&=\delta_{ab}\,,&
  \osca_a|0\rangle&=0\,,\\\label{dual}
  \bar{J}_{ab}&=-\oscbb_b \oscb_a&
  &\text{with}&
  [\oscb_a,\oscbb_b]&=\delta_{ab}\,,&
  \oscb_a|\bar 0\rangle&=0\,,&
\end{align}
where the fundamental indices $a,b$ take the values $1,2$. We call
\eqref{symm} a {\em symmetric realization} and \eqref{dual} a {\em
  dual realization}. The generators $J_{ab}$ and $\bar{J}_{ab}$ act on
the states obtained by applying the creation operators
$\oscab_a$ and $\oscbb_a$ to their respective Fock vacua $|0\rangle$ and  $|\bar 0\rangle$. The infinite-dimensional vector
space spanned by these states decomposes into finite-dimensional representation spaces $V_s$
and $\bar V_s$ of homogeneous polynomials of degree $s$ in the creation operators. For each
degree there is a highest weight state and the representation is
labeled by the positive integer $s$ which is an eigenvalue of one of
the Cartan elements,
\begin{align}\label{hwssym}
  |\text{hws}\rangle &= (\oscab_1)^s|0\rangle\,,&
  J_{aa}|\text{hws}\rangle &=s\,\delta_{a1} |\text{hws}\rangle\,,\\
  \label{hwsdual}
  \overline{|\text{hws}\rangle} &= (\oscbb_2)^s|\bar 0\rangle\,,&
  \bar{J}_{aa}\overline{|\text{hws}\rangle}&=-s\,\delta_{a2} \overline{|\text{hws}\rangle}\,.
\end{align}
It is sometimes convenient to notationally hide the difference between the two types of oscillators $\osca$ and $\oscb$, and to instead use only one type $\oscw$ satisfying
\begin{equation}
[\oscw_a,\oscwb_b]=\delta_{ab}\,.
\end{equation}
This is just a relabeling which looks as follows:
\begin{align}
  \oscab_a&\leftrightarrow\oscwb_a\,,&
  \osca_a&\leftrightarrow\oscw_a\,,&
  \oscw_a|0\rangle&=0\,,\label{eq:arelabeling}\\
  \oscbb_a&\leftrightarrow -\oscw_a\,,&
  \oscb_a&\leftrightarrow\oscwb_a\,,&
  \oscwb_a|\bar 0\rangle&=0\label{eq:brelabeling}\,.
\end{align}
Note that we are not spelling out any conjugation properties of our oscillators, nor computing any norms, therefore there is no problem with negative norm states from \eqref{eq:brelabeling}.

Written in terms of these variables the generators and highest weight
states of $V_s$ and $\bar V_s$ become, respectively,
\begin{align}
  \label{eq:realdiffop}
  J_{ab}&=\oscwb_a\oscw_b\,,&
  |\text{hws}\rangle &=( +\oscwb_1)^s |0\rangle\,,\\
  \label{eq:realdiffopbar}
  \bar{J}_{ab}&=\oscw_b \oscwb_a\,,&
  \overline{|\text{hws}\rangle} &= (-\oscw_2)^s|\bar 0\rangle\,.
\end{align}
In the following we will use both notations, as is convenient.

The space $V=V_1\otimes \ldots\otimes V_n$ of \eqref{invariance} can
then be built out of factors $V_i$ which are of the type 
$V_{s_i}:={\rm span}\{\oscwb^{s_i}|0\rangle\}$ or 
${\bar V}_{s_i}:={\rm span}\{(-\oscw)^{s_i}|\bar 0\rangle\}$. 
We may think of $V$ as the quantum space of a compact spin
chain. The monodromy matrix $M(u)$ of this spin chain is defined on
$V_\square\otimes V$, where $V_\square$ denotes an auxiliary
space in the fundamental representation $\square$ of $\mathfrak{gl}(2)$. The
monodromy matrix can be written with the help of Lax operators $L(u,v)$ and $\bar{L}(u,v)$ describing the ``interaction'' of the auxiliary space with, respectively, the spaces
$V_s$ and $\bar V_s$. We use similar Lax operators as in
\cite{Frassek:2013xza}. For the symmetric representations we take
\begin{align}\label{Lax}
  L(u,v)= 1+(u-v)^{-1}\sum_{a,b}e_{ab}\,\oscab_b\osca_a=   
\,\,  \raisebox{-0.55\height}{\scalebox{0.35}{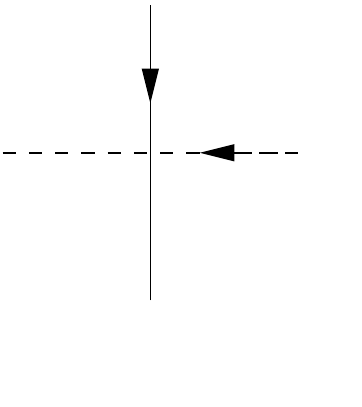}}\qquad\,,
\end{align}
while for dual ones
\begin{align}\label{Laxbar}
  \bar L(u,v)= 1-(u-v-1)^{-1}\sum_{a,b}e_{ab}\,\oscbb_a\oscb_b\,.
\end{align}
In both cases the elementary matrices $e_{ab}$ with matrix elements
$(e_{ab})_{cd}=\delta_{ac}\delta_{bd}$ act on the auxiliary
space. Compared to \cite{Frassek:2013xza} we dropped a non-trivial
normalization factor of the Lax operators and we introduced a shift of
the parameter $v$ in \eqref{Laxbar}. This shift allows us to express
both types of Lax operators in terms of
\begin{align}\label{curlyLax}
  \mathcal{L}(u,v)=u-v+\sum_{a,b}e_{ab}\,\oscwb_{b}\oscw_a\,.
\end{align}
Using \eqref{eq:arelabeling} and \eqref{eq:brelabeling},
respectively, we obtain
\begin{align}
  \label{laxcurlylax}
  L(u,v)&=(u-v)^{-1}\mathcal{L}(u,v)\,,\\
  \label{laxbarcurlylax}
  \bar L(u,v)&=(u-v-1)^{-1}\mathcal{L}(u,v)\,.
\end{align}
In order to render our discussion clearer, we take a spin chain with a
very particular quantum space. The first $k$ sites are represented with the
use of the dual realization \eqref{dual}, and the last $n-k$ sites with
the symmetric realization \eqref{symm}, i.e.\
\begin{align}
  \label{qspace}
  V=\bar V_{s_1}\otimes\cdots\otimes\bar V_{s_k}\otimes 
  V_{s_{k+1}}\otimes \cdots\otimes V_{s_{n}}\,.
\end{align}
Finally, the monodromy matrix reads
\begin{align}\label{monodromy}
  M(u)&=
  \bar{L}_1(u,v_1)\ldots \bar{L}_k(u,v_{k})
  L_{k+1}(u,v_{k+1})\ldots L_n(u,v_n)\\
  &
  \begin{aligned}
    \scalebox{0.60}{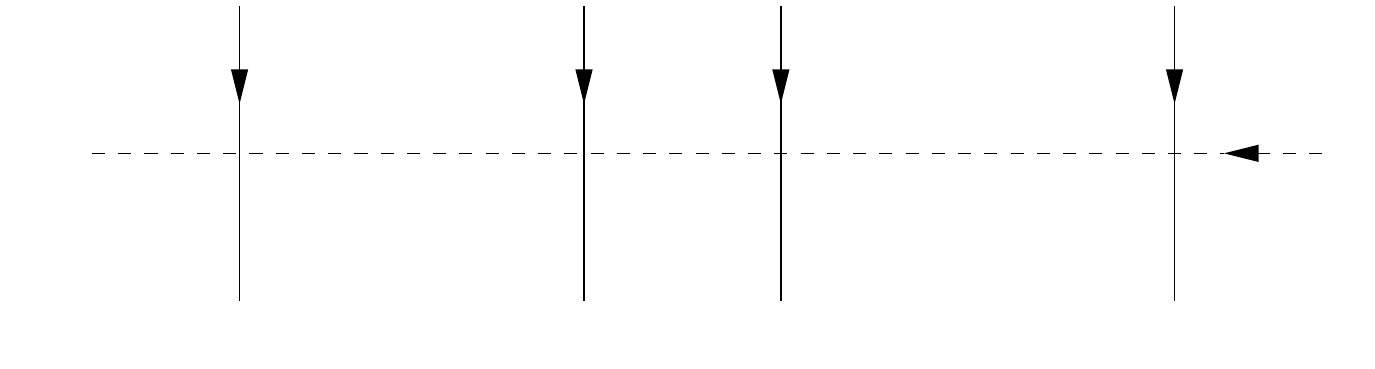}
  \end{aligned}\qquad\,.
\end{align}
Note that the monodromy $M(u)$ depends only on the spectral parameters $u$ and $v_i$ as well as on $n$ and $k$, but not on the representation labels $s_i$. As we did already in \eqref{Lax}, we have nevertheless attached these labels in the graphical depiction, to indicate the nature of the quantum space the monodromy is acting on. This monodromy provides a realization of the Yangian
$Y(\mathfrak{gl}(2))$. Each Lax operator itself is an evaluation
realization of the Yangian with evaluation parameter $v_i$, which is
called an inhomogeneity in spin chain language.

\subsection{Quantum Inverse Scattering Method for Yangian Invariants}

It was shown in \cite{Frassek:2013xza} that one can construct
eigenvectors $|\Psi \rangle$ satisfying \eqref{invariance} using the
Quantum Inverse Scattering Method (QISM), and in particular apply the
Algebraic Bethe Ansatz technique. Most details of this construction
were given for the simplest case of compact $\mathfrak{gl}(2)$
representations. In this case one writes the monodromy matrix
\eqref{monodromy} explicitly as a matrix acting in the fundamental
auxiliary space,
\begin{equation}
  M(u)=
  \begin{pmatrix}
    A(u)&B(u)\\
    C(u)&D(u)\\
  \end{pmatrix}.
\end{equation}
The operators $A(u), B(u),C(u)$ and $D(u)$ act only on the quantum
space $V$. The standard procedure is then to construct a ``reference''
state $|\Omega\rangle$ satisfying
\begin{align}
  C(u)|\Omega\rangle=0\,,\qquad 
  A(u)|\Omega\rangle=\alpha(u)|\Omega\rangle\,,\qquad
  D(u)|\Omega\rangle=\delta(u)|\Omega\rangle\,,
\end{align}
where $\alpha(u)$ and $\delta(u)$ are scalar functions depending on
the representations labels $s_i$ and the inhomogeneities $v_i$ of the
monodromy. The reference state $|\Omega\rangle$ is realized as the
tensor product of the highest weight states at each spin chain
site. As already pointed out in \eqref{qspace}, we focus on the case
where the first $k$ highest weight states are of the form
\eqref{hwsdual} and the remaining $n-k$ are as given in
\eqref{hwssym},
\begin{align}
  |\Omega \rangle=
  \bigotimes_{i=1}^k \overline{|\text{hws}\rangle}_i\bigotimes_{i=k+1}^n |\text{hws}\rangle_i=
  \prod_{i=1}^k(\oscbb_2^i)^{s_i}\prod_{i=k+1}^n (\oscab_1^i)^{s_i}\,
  |\mathbf{0}\rangle
\end{align}
with
\begin{align}\label{multifock}
  |\mathbf{0}\rangle=
    \underbrace{|\bar 0\rangle \otimes \ldots \otimes |\bar 0\rangle}_k \otimes 
  \underbrace{|0\rangle \otimes \ldots \otimes |0\rangle}_{n-k}\,.
\end{align}
In order to construct Yangian invariants one proceeds to define the
Bethe vectors
\begin{equation}
|\Psi\rangle_{n,k}=B(u_1)\ldots B(u_K)|\Omega\rangle\,.
\end{equation} 
The Bethe roots $u_j$ and the parameters $s_i$ and $v_i$ of the
monodromy have to satisfy the first order Baxter equations
\begin{gather}
  \label{baxterq}
  \frac{Q(u)}{Q(u+1)}=\delta(u)\,\qquad\text{with}\qquad
  Q(u)=\prod_{j=1}^K (u-u_j)\,,\\
  \label{baxterad}
  \alpha(u)\delta(u-1)=1\,.
\end{gather}
These equations impose stronger conditions than the usual Bethe
equations and they guarantee the Yangian invariance of the Bethe
vector. For the monodromy \eqref{monodromy} the functions $\alpha(u)$
and $\delta(u)$ can be worked out explicitly. This turns
\eqref{baxterq} into an equation\footnote{Equations \eqref{1stBE} and
  \eqref{2ndbe} are slightly different from the ones in
  \cite{Frassek:2013xza}. This difference originates from the shift of
  the inhomogeneity in the dual Lax operator \eqref{Laxbar}.}
determining the Bethe roots $u_j$
\begin{equation}\label{1stBE}
\frac{Q(u)}{Q(u+1)}=\prod_{i=1}^k \frac{u-v_i-s_i-1}{u-v_i-1}\,.
\end{equation}
In addition, \eqref{baxterad} becomes
\begin{equation}\label{2ndbe}
\prod_{i=1}^k \frac{u-v_i-s_i-2}{u-v_i-2}\prod_{i=k+1}^n \frac{u-v_i+s_i}{u-v_i}=1\,,
\end{equation}
which constrains the representation labels $s_i$ and inhomogeneities
$v_i$. 

Equations \eqref{1stBE} and \eqref{2ndbe} were solved explicitly in
\cite{Frassek:2013xza} for some sample invariants. The corresponding
Bethe vectors were evaluated for small integer values of the
representation labels $s_i$. Up to overall normalization factors this
led to the invariants
\begin{align}\label{psi21}
  |\Psi\rangle_{2,1}&=(\oscbb^1 \cdot \oscab^2)^{s_2} |\mathbf{0}\rangle\,,\\\label{psi31}
  |\Psi\rangle_{3,1}&=(\oscbb^1 \cdot \oscab^2)^{s_2}(\oscbb^1 \cdot \oscab^3)^{s_3} |\mathbf{0}\rangle\,,\\\label{psi32}
  |\Psi\rangle_{3,2}&=(\oscbb^1 \cdot \oscab^3)^{s_1}(\oscbb^2 \cdot \oscab^3)^{s_2} |\mathbf{0}\rangle\,,
  \\\label{psi42}
  |\Psi\rangle_{4,2}&=\sum_{k=0}^{\infty}\frac{1}{(s_1-k)!(s_2-k)!k!\Gamma(z-s_1+k+1)}\\
  \nonumber &\hspace{1cm}\cdot(\oscbb^1 \cdot\oscab^3)^{s_1-k}(\oscbb^2 \cdot\oscab^4)^{s_2-k}(\oscbb^2 \cdot\oscab^3)^{k}(\oscbb^1 \cdot\oscab^4)^{k}|\mathbf{0}\rangle\,,
\end{align}
where $\oscbb^i\cdot \oscab^i=\sum_a \oscbb_a^i
\oscab_a^i$. Surprisingly, perhaps, even though one acts in general
with a large number of suitable operators $B(u_j)$ on the reference
state $|\Omega\rangle$, the final result looks very simple for the
first few invariants $|\Psi\rangle_{2,1}$, $|\Psi\rangle_{3,1}$ and
$|\Psi\rangle_{3,2}$. As for the considerably more involved ``harmonic R-matrix'' $|\Psi\rangle_{4,2}$, note the following two features. Firstly, for finite dimensional representations, the sum in \eqref{psi42} is actually finite, of course. Secondly, a complex spectral parameter $z$ appears.

The expressions \eqref{psi21}-\eqref{psi32} are reminiscent of the
ones found in \cite{Chicherin:2013ora}, where the authors studied a
condition for Yangian invariance\footnote{Up to a very non-trivial
  normalization of the monodromy, which however will not play a role
  in this paper.} similar to \eqref{invariance}. Even though the method
used there differs from the standard Algebraic Bethe Ansatz approach
explained above, the idea bears many similarities. The construction in
\cite{Chicherin:2013ora} is valid for any algebra $\mathfrak{gl}(N|M)$.
Let us explain it here in a few steps, restricting to the
$\mathfrak{gl}(2)$ case for simplicity. Now it is convenient to use \eqref{eq:arelabeling} and
\eqref{eq:brelabeling} to express the $\osca$, $\oscb$ oscillators in terms of the
$\oscw$ oscillators. As opposed to the reference state $|\Omega\rangle$, the authors of
\cite{Chicherin:2013ora} start from the Fock vacuum $|\mathbf{0}\rangle$  in the quantum space, defined in \eqref{multifock}, which corresponds to a trivial singlet representation. They then look for
Yangian invariants of the form\footnote{We changed the notation from
  $R_{ij}(u)$ to $\mathcal{B}_{ij}(u)$ compared to
  \cite{Chicherin:2013ora}, and furthermore changed the normalization.}
\begin{equation}\label{eigenvecChicherin}
|\Psi\rangle=\mathcal{B}_{i_{1} j_{1}}(\bar{u}_{1})\ldots\mathcal{B}_{i_{P} j_{P}}(\bar{u}_{P})|\mathbf{0}\rangle\,,
\end{equation} 
with
\begin{equation}\label{operatorB}
\mathcal{B}_{i j}(u)=(-\oscwb^j \cdot \oscw^i)^u
\end{equation}
and $i,j=1,\ldots n$. As an example let us consider the invariants in
\eqref{psi21}, \eqref{psi31}, \eqref{psi32} and notice that
\begin{align}\label{Psi21C}
|\Psi\rangle_{2,1}&\propto\mathcal{B}_{12}(s_2)|\mathbf{0}\rangle\,,\\\label{Psi31C}
|\Psi\rangle_{3,1}&\propto\mathcal{B}_{12}(s_2)\mathcal{B}_{13}(s_3)|\mathbf{0}\rangle\,,\\\label{Psi32C}
|\Psi\rangle_{3,2}&\propto\mathcal{B}_{13}(s_1)\mathcal{B}_{23}(s_2)|\mathbf{0}\rangle\,.
\end{align} 
At first, it seems impossible to also write down the four-point
invariant in such a simple form since it is given in \eqref{psi42} as
a complicated sum. In addition, it depends on the complex parameter $z$. However, using the formal algebraic commutation relations for the oscillators, one easily proves
\begin{equation}\label{42fromB}
  |\Psi\rangle_{4,2} \propto 
  \mathcal{B}_{12}(z)\mathcal{B}_{23}(s_1)\mathcal{B}_{12}(s_1-z)\mathcal{B}_{24}(s_2) |\mathbf{0}\rangle\,,
\end{equation}
or more explicitly
\begin{equation}\label{rewrite42}
  |\Psi\rangle_{4,2}=
  \frac{1}{\Gamma(s_1+1)\Gamma(s_2+1)\Gamma(z+1)}
  (\oscbb^1 \cdot\oscb^2)^{z}(\oscbb^2 \cdot\oscab^3)^{s_1}
  (\oscbb^1 \cdot\oscb^2)^{s_1-z}(\oscbb^2 \cdot\oscab^4)^{s_2}|\mathbf{0}\rangle\,.
\end{equation}
The careful reader might be a bit puzzled here since in
\eqref{42fromB} and \eqref{rewrite42} the spectral parameter $z$ is a
complex number. Let us therefore give some details on the purely
algebraic derivation of \eqref{psi42} from \eqref{rewrite42}. We
interpret the action of the two rightmost operators as
\begin{align}
  (\oscbb^1 \cdot\oscb^2)^{s_1-z}(\oscbb^2 \cdot\oscab^4)^{s_2}|\mathbf{0}\rangle=
  \frac{\Gamma(s_2+1)}{\Gamma(s_2-s_1-z+1)}
  (\oscbb^2 \cdot\oscab^4)^{s_2-s_1+z}(\oscbb^1 \cdot\oscab^4)^{s_1-z}|\mathbf{0}\rangle\,.
\end{align}
Then the fourth operator $(\oscbb^1 \cdot\oscb^2)^{z}$ produces the
sum of terms in \eqref{psi42} by means of the generalized Leibniz
rule. Finally, all complex powers of oscillators disappear and we
obtain the invariant \eqref{psi42} for compact representations. This
illustrates that we should not require in \eqref{operatorB} the
variable $u$ to be a non-negative integer; we take it as a general
complex number $u \in \mathbb{C}$.

In \cite{Chicherin:2013ora} the authors studied a few more examples
beyond \eqref{Psi21C}, \eqref{Psi31C} and \eqref{Psi32C}. However, no
general construction for arbitrary $n$ and $k$ was proposed. In particular,
no classification of such invariants was provided. In practical terms,
we would like to find a prescription which tells us which
$(i_p,j_p)$ and $\bar{u}_p$ in \eqref{eigenvecChicherin} have to be
taken in order to obtain the Yangian invariant $|\Psi\rangle_{n,k}$
for a given $n$ and $k$. We will now fill this gap. It
will turn out that there are many distinct solutions for fixed $n$ and
$k$, but that all of them are in one-to-one correspondence with permutations.

Before we proceed to this general construction let us come back for a
moment to the first order Baxter equation \eqref{baxterad} spelled out
in \eqref{2ndbe}. There exists a redefinition of the inhomogeneities
which turns this equation into a much simpler form. Let us introduce
\begin{align}
  \label{vprime}
  v_i'=
  \begin{cases}
    v_i+\tfrac{s_i}{2}+2&\quad\text{for}\qquad i=1,\ldots,k\,,\\
    v_i-\tfrac{s_i}{2}&\quad\text{for}\qquad i=k+1,\ldots,n\,.\\
  \end{cases}
\end{align} 
In addition we define
\begin{align}
  \label{symduals}
  \mathfrak{s}_i=
  \begin{cases}
    -s_i&\quad\text{for}\qquad i=1,\ldots,k\,,\\
    s_i&\quad\text{for}\qquad i=k+1,\ldots,n\,,\\
  \end{cases}
\end{align}
and
\begin{align}
  v_i^{\pm}=v_i'\pm \sfrac{\mathfrak{s}_i}{2}\,.
\end{align}
In term of these variables \eqref{2ndbe} reads
\begin{equation}
\prod_{i=1}^n (u-v_i^+)=\prod_{i=1}^n (u-v_i^-)\,.
\end{equation}
Both sides of this equation are polynomials in $u$ of degree
$n$. Since the set of roots is unique for a given polynomial we
conclude that for each $v_i^+$ there exists $j$ such that
$v_{i}^+=v_{j}^-$. Altogether, it means that with each solution to
\eqref{2ndbe} we can associate a permutation $\sigma$ such that
\begin{equation}\label{vpvm}
v_{\sigma(i)}^+=v_i^-\,.
\end{equation}
This provides a systematic construction of the solutions of
\eqref{baxterad}, a problem proposed in \cite{Frassek:2013xza}. We
restrict our representations from now on and take generic values for
the $s_i$, namely $s_i\neq 0$, then $\sigma(i)\neq i$, i.e.~$\sigma$
does not have fixed points.

Even though the above analysis was done for the case of
$\mathfrak{gl}(2)$ we claim that it continues to be valid for any
realization of the $\mathfrak{gl}(N|M)$ algebra in terms of a single
family of oscillators. In particular this includes the
$\mathfrak{gl}(4|4)$ case relevant for the $\mathcal{N}=4$ SYM
amplitudes. It is not an accident that on-shell diagrams are also
parametrized by permutations. As was shown in \cite{Ferro:2013dga}
each on-shell diagram can be deformed by allowing non-physical
particle helicities. While a generic deformation of the amplitude is
not Yangian invariant, there exists a simple criterion enforcing
it. It is sufficient to restrict to deformations such that all
possible cluster transformations leave the appropriate integration
measure invariant, see \cite{Ferro:2013dga} for details. These
constraints were studied in details in \cite{Beisert:2014qba} and it was shown
there that they can be naturally recast in the form \eqref{vpvm}. We
will show that all such deformations can be obtained from
\eqref{eigenvecChicherin} with an appropriate set of indices
$(i_p,j_p)$ and parameters $\bar{u}_p$, leading to a complete
classification of Yangian invariants relevant for the tree-level
amplitudes.

\subsection{General Construction of Yangian Invariants}
\label{genconstr}

In this section we will construct a Yangian invariant
$|\Psi\rangle_\sigma$ for each permutation $\sigma$ using the
operators $\mathcal{B}_{ij}(u)$ introduced in \eqref{operatorB}. The procedure is closely related to the combinatorics of permutations and scattering amplitudes in chapter 2 of \cite{ArkaniHamed:2012nw}.

Let $\sigma$ be a permutation of $n$ elements and $k$ be the number of
elements $i$ with $\sigma(i)<i$. We decompose $\sigma$ into
transpositions $T_p=(i_p,j_p)$, which exchange the two
elements $i_p<j_p$
\begin{align}
  \label{decomposition}
  \sigma
  =T_{P}\circ\ldots\circ T_2\circ T_{1}
  =(i_{P},j_{P})\cdots(i_2,j_2)(i_{1},j_{1})\,.
\end{align} 
This (non-unique) decomposition is assumed to be \emph{minimal}, meaning that there
exists no other decomposition of $\sigma$ into a smaller number of
transpositions. In addition, the transpositions $T_p=(i_p,j_p)$ are
required to be \emph{adjacent}, cf.\ \cite{ArkaniHamed:2012nw}, in the sense that
\begin{align}
  \label{adjacent}
  i_q,j_q\notin\{i_p+1,\ldots, j_p-1\}\qquad\text{for}\qquad q>p\,.
\end{align}
Note that this ``generalized adjacency'' of the indices $i_p$, $j_p$ of the transposition $T_p$ means that the condition $i_p+1=j_p$ is relaxed to $i_p+1<j_p$, iff all transpositions $T_q=(i_q,j_q)$ applied after $T_p$ are restricted by \eqref{adjacent}.
For practical purposes, the Mathematica program in \cite{Bourjaily:2012gy} may be used to obtain
such a decomposition for a given permutation $\sigma$.

\begin{figure}[ht]
\begin{center}
\scalebox{0.55}{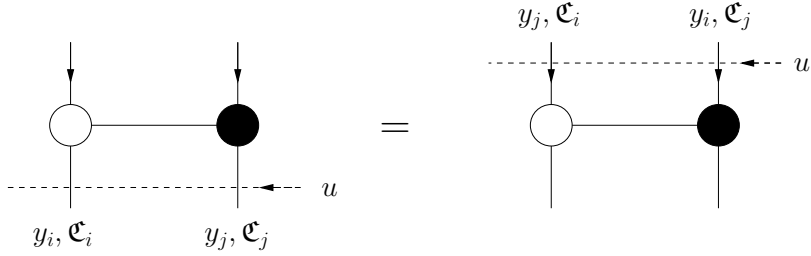}
\end{center}
\caption{Fundamental relation \eqref{YBElike} for the construction of
  Yangian invariants using the intertwining operators
  $\mathcal{B}_{ij}(y_i-y_j)$.}
\label{Fig:YBE}
\end{figure}

We claim that a Yangian invariant is now constructible from the ansatz
\eqref{eigenvecChicherin}, which reads
\begin{align}
  \label{ansatzyi}
  |\Psi\rangle_\sigma=
  \mathcal{B}_{i_1 j_1}(\bar{u}_{1})\cdots
  \mathcal{B}_{i_P j_P}(\bar{u}_{P})|\mathbf{0}\rangle\,.
\end{align}
Here the indices of the operators are precisely the arguments of the
transpositions in \eqref{decomposition}, and the number $k$ defined
above fixes the vacuum $|\mathbf{0}\rangle$ to be
\eqref{multifock}. The parameters $\bar{u}_p$, akin to Bethe roots in
the Algebraic Bethe Ansatz, will be given below. Let us briefly
outline the idea of how to show the Yangian invariance of this
ansatz. The main tool is the ``intertwining'' relation
\begin{align}
  \label{YBElike}
    \mathcal{L}_{i}(u,y_i+\mathfrak{C}_i)
    \mathcal{L}_{j}(u,y_j+\mathfrak{C}_j)
    \mathcal{B}_{ij}(y_i-y_j)
    =\mathcal{B}_{ij}(y_i-y_j)
    \mathcal{L}_{i}(u,y_j+\mathfrak{C}_i)
    \mathcal{L}_{j}(u,y_i+\mathfrak{C}_j)\,,
\end{align}
with $\mathfrak{C}_i=\oscwb^i\cdot\oscw^i$ being the number operator
of the oscillators at site $i$. This equation is easily verified by a
direct computation, and may be depicted as in Figure~\ref{Fig:YBE}. It
is similar to, but distinct from the standard form of the Yang-Baxter
equation: Note that the indices on the Lax operators $\mathcal{L}_{i}$
and the number operators $\mathfrak{C}_i$ are not permuted, while the
ones on the inhomogeneities $y_i$ are. A closely related equation was
introduced in \cite{Chicherin:2013ora,Kirschner:2013ila}, see also
appendix~\ref{interwining-relation}.
One now expresses the monodromy $M(u)$ given in \eqref{monodromy} entirely in terms of Lax
operators $\mathcal{L}_{i}(u,v_i)$ found in \eqref{curlyLax}, and then acts with $M(u)$ on the ansatz \eqref{ansatzyi}. Then one uses \eqref{YBElike} to commute all operators
$\mathcal{B}_{i_pj_p}(\bar{u}_p)$ inside $|\Psi\rangle_\sigma$ to the
left side of the monodromy $M(u)$. This constrains the
parameters $\bar{u}_p$ in terms of the inhomogeneities $v_i$ of the
monodromy. It also modifies the distribution of inhomogeneities of the monodromy. However, one notices that this modified monodromy acts diagonally on the
vacuum $|\mathbf{0}\rangle$. This shows that the ansatz \eqref{ansatzyi}
satisfies the invariance condition \eqref{invariance} up to a factor.
The latter conveniently turns out to be unity in our conventions.

To implement this procedure we replace the inhomogeneities $v_i$ of
the monodromy by new variables\footnote{These variables are related to those in
  \eqref{vprime} by $y_i=v_i'-\tfrac{\mathfrak{s}_i}{2}$.} $y_i$ defined in terms of the
representation labels $\mathfrak{s}_i$, cf.\ \eqref{symduals}
\begin{align}
  \label{redefinh}
  v_i|\Psi\rangle_\sigma=
  (y_i+\mathfrak{C}_i)|\Psi\rangle_\sigma=
  \begin{cases}
    (y_i+\mathfrak{s}_i-N+M)|\Psi\rangle_\sigma&\qquad\text{for}\qquad i=1,\ldots,k\,,\\
    (y_i+\mathfrak{s}_i)|\Psi\rangle_\sigma&\qquad\text{for}\qquad i=k+1,\ldots,n\,.\\
  \end{cases}
\end{align}
For the convenience of the reader, here and in the formulas in the rest of this section we have already given the correct expressions for the algebra $\mathfrak{gl}(N|M)$; to specialize back to $\mathfrak{gl}(2)$ just put $N=2$, $M=0$.
The action of the monodromy on the ansatz \eqref{ansatzyi} in these variables reads
\begin{align}
  \label{eq:action mono}
  M(u)|\Psi\rangle_\sigma=
  \prod_{i=1}^k\frac{1}{u-v_i-1}\prod_{i=k+1}^n\frac{1}{u-v_i}
  \mathcal{L}_{1}(u,y_1+\mathfrak{C}_1)\cdots
  \mathcal{L}_{n}(u,y_n+\mathfrak{C}_n)
  |\Psi\rangle_\sigma\,.
\end{align}
The arguments of the Lax operators are already in the form of
\eqref{YBElike}. To apply \eqref{YBElike} we also have to specify the
variables $\bar{u}_p$ in \eqref{ansatzyi} in terms of the $y_i$. For
this purpose we introduced the permutations
\begin{align}
  \label{chainperm}
  \tau_p=\tau_{p-1}\circ(i_p,j_p)=(i_1,j_1)\cdots(i_p,j_p)
\end{align}
for $p=1,\ldots,P$. As we will see shortly, the correct choice of the $\bar{u}_p$ is
\begin{equation}
  \label{baru}
  \bar{u}_p=y_{\tau_{p}(j_p)}-y_{\tau_{p}(i_p)}\,.
\end{equation} 
Let us explain how to use \eqref{YBElike} to commute the operators
$\mathcal{B}_{i_pj_p}(\bar{u}_p)$ inside $|\Psi\rangle_\sigma$ to the
left side of the monodromy $M(u)$. For a moment we disregard the
arguments $\bar{u}_p$. If the indices of an operator
$\mathcal{B}_{i_pj_p}$ are adjacent, $i_p+1=j_p$, we can directly
apply \eqref{YBElike} to move $\mathcal{B}_{i_pj_p}$ left of the Lax
operators $\mathcal{L}_{i_p}\mathcal{L}_{i_{p}+1}$ and then on through the entire monodromy. 
However, if $i_p+1>j_p$, this is not immediately possible because in the monodromy
there are some Lax operators in between $\mathcal{L}_{i_p}$ and $\mathcal{L}_{j_p}$. In this case the
decomposition into adjacent transpositions guarantees, cf.\
\eqref{adjacent}, that all operators $\mathcal{B}_{i_qj_q}$ to the right of
$\mathcal{B}_{i_pj_p}$ do not act on the spaces labeled
$i_p+1,\ldots,j_p-1$. Hence, the corresponding Lax operators
$\mathcal{L}_{i_p+1}$, \ldots, $\mathcal{L}_{j_p-1}$ in the monodromy act directly on the vacuum
$|\mathbf{0}\rangle$. For any Lax operator acting on the vacuum we have
\begin{align}
  \label{laxonvac}
  \mathcal{L}_{i}(u,y_j+\mathfrak{C}_i)|\mathbf{0}\rangle=
  \begin{cases}
    (u-y_j+N-M-1)|\mathbf{0}\rangle&\qquad\text{for}\qquad i=1,\ldots,k\,,\\
    (u-y_j)|\mathbf{0}\rangle&\qquad\text{for}\qquad i=k+1,\ldots,n\,.
  \end{cases}
\end{align}
Consequently, the Lax operators between $\mathcal{L}_{i_p}$ and $\mathcal{L}_{j_p}$ effectively disappear
from the monodromy. This means we can apply \eqref{YBElike} to commute
the operators $\mathcal{B}_{i_pj_p}$ past the monodromy also in the
case $i_p+1>j_p$.

Finally, we give some details on the commutation of all operators
$\mathcal{B}_{i_pj_p}(\bar{u}_p)$ inside $|\Psi\rangle_\sigma$ to the
left of the monodromy $M(u)$. For this computation we introduce
\begin{equation}
  \mathcal{M}(u;y_1,\ldots,y_n)=
  \mathcal{L}_{1}(u,y_1+\mathfrak{C}_1)\,
  \mathcal{L}_{2}(u,y_2+\mathfrak{C}_2)\cdots
  \mathcal{L}_{n}(u,y_n+\mathfrak{C}_n)\,.
\end{equation}
Using \eqref{YBElike}, \eqref{baru} and applying \eqref{laxonvac} once
the whole monodromy hits the vacuum yields
\begin{align}
  &M(u)|\Psi\rangle_\sigma=
  \prod_{i=1}^k\frac{1}{u-v_i-1}\prod_{i=k+1}^n\frac{1}{u-v_i}
  \mathcal{M}(u;y_1,\ldots,y_n)
  \mathcal{B}_{i_{1}j_{1}}(\bar{u}_{1})\cdots
  \mathcal{B}_{i_{P}j_{P}}(\bar{u}_{P})|\mathbf{0}\rangle\\
  &=
  \prod_{i=1}^k\frac{1}{u-v_i-1}\prod_{i=k+1}^n\frac{1}{u-v_i}
  \mathcal{B}_{i_{1}j_{1}}(\bar{u}_{1})
  \mathcal{M}(u;\ldots,y_{j_{1}},\ldots,y_{i_{1}},\ldots)
  \mathcal{B}_{i_{2}j_{2}}(\bar{u}_{2})\cdots
  \mathcal{B}_{i_{P}j_{P}}(\bar{u}_{P})|\mathbf{0}\rangle\\
  &=
  \prod_{i=1}^k\frac{1}{u-v_i-1}\prod_{i=k+1}^n\frac{1}{u-v_i}
  \mathcal{B}_{i_{1}j_{1}}(\bar{u}_{1})
  \mathcal{M}(u;y_{\tau_{1}(1)},\ldots,y_{\tau_{1}(n)})
  \mathcal{B}_{i_{2}j_{2}}(\bar{u}_{2})\cdots
  \mathcal{B}_{i_{P}j_{P}}(\bar{u}_{P})|\mathbf{0}\rangle\,\\
  &=
  \prod_{i=1}^k\frac{1}{u-v_i-1}\prod_{i=k+1}^n\frac{1}{u-v_i}
  \mathcal{B}_{i_{1}j_{1}}(\bar{u}_{1})
  \mathcal{B}_{i_{2}j_{2}}(\bar{u}_{2})\cdots
  \mathcal{B}_{i_{P}j_{P}}(\bar{u}_{P})
  \mathcal{M}(u;y_{\tau_{P}(1)},\ldots,y_{\tau_{P}(n)})|\mathbf{0}\rangle\\
  &=
  \prod_{i=1}^k\frac{u-y_{\tau_{P}(i)}+N-M-1}{u-v_i-1}
  \prod_{i=k+1}^n\frac{u-y_{\tau_{P}(i)}}{u-v_i}
  \mathcal{B}_{i_{1}j_{1}}(\bar{u}_{1})
  \mathcal{B}_{i_{2}j_{2}}(\bar{u}_{2})\cdots
  \mathcal{B}_{i_{P}j_{P}}(\bar{u}_{P})|\mathbf{0}\rangle\\
  &=
  \prod_{i=1}^k\frac{u-y_{\tau_{P}(i)}+N-M-1}{u-v_i-1}
  \prod_{i=k+1}^n\frac{u-y_{\tau_{P}(i)}}{u-v_i}
  |\Psi\rangle_\sigma\,.\label{evcalc}
\end{align}
This proves that the ansatz for $|\Psi\rangle_\sigma$ given in
\eqref{ansatzyi} with $\bar{u}_p$ specified in \eqref{baru} satisfies
the Yangian invariance condition \eqref{invariance} up to a
factor. Let us now argue that this factor is equal to $1$. One may compute the representation labels $\mathfrak{s}_i$ of the vector $|\Psi\rangle_\sigma$, leading to
\begin{align}
  \label{reblabelinv}
  y_{\sigma(i)}+\mathfrak{s}_{\sigma(i)}=y_i\,.
\end{align}
Together with the variable redefinition \eqref{redefinh} and
$\tau_P=\sigma^{-1}$ this turns the scalar factor in \eqref{evcalc}
into $1$, which shows the Yangian invariance of
$|\Psi\rangle_\sigma$. Notice that defining $v_i^+=y_i+\mathfrak{s}_i$
and $v_i^-=y_i$ turns \eqref{reblabelinv} into \eqref{vpvm}.


\section{Generalization to Superalgebras and Scattering Amplitudes}
\label{Sec:Amplitudes}

The general construction of Yangian invariants in the previous section
easily generalizes to compact representations of the superalgebra
$\mathfrak{gl}(N|M)$ that can be realized in terms of a single
oscillator family. Moreover, we will argue that it also applies to the
non-compact representations of $\mathfrak{gl}(4|4)$ relevant to
tree-level scattering amplitudes of $\mathcal{N}=4$ SYM.

Let us first focus on the compact representations of
$\mathfrak{gl}(N|M)$. We therefore replace the oscillators of
section~\ref{Sec:Construction} by superoscillators. Generalizing
\eqref{symm} and \eqref{dual}, we obtain two families of
realizations with generators
\begin{align}
  \label{supergensym}
  J_{ab}&=+\oscab_a \osca_b&
  &\text{with}&
  [\osca_a,\oscab_b\}&=\delta_{ab}\,,&
  \osca_a|0\rangle&=0\,,\\
  \label{supergendual}
  \bar{J}_{ab}&=-(-1)^{b+ab}\oscbb_b \oscb_a&
  &\text{with}&
  [\oscb_a,\oscbb_b\}&=\delta_{ab}\,,&
  \oscb_a|\bar{0}\rangle&=0\,,&
\end{align}
where the oscillators labeled by $a,b=1,\ldots,N$ are bosonic and
those with $a,b=N+1,\ldots,N+M$ are fermionic, and the exponent of $-1$ is
to be understood as the degree of the corresponding index. As in the
purely bosonic case it is convenient to relabel the oscillators
by introducing yet another oscillator family $\oscw$ as in \eqref{eq:arelabeling} and \eqref{eq:brelabeling},
\begin{align}
  \oscab_a&\leftrightarrow\oscwb_a\,,&
  \osca_a&\leftrightarrow\oscw_a\,,&
  \oscw_a|0\rangle&=0\,,\\
  \oscbb_a&\leftrightarrow -(-1)^a\,\oscw_a\,,&
  \oscb_a&\leftrightarrow\oscwb_a\,,&
  \oscwb_a|\bar 0\rangle&=0\,.
\end{align}
The Lax operator \eqref{curlyLax} then generalizes to a graded one,
\begin{align}
  \mathcal{L}(u,v)=u-v+\sum_{a,b}(-1)^b e_{ab}\,\oscwb_b\oscw_a\,,
\end{align}
whence the $\mathfrak{gl}(N|M)$ versions of $L(u,v)$ and
$\bar{L}(u,v)$ may be deduced by imposing \eqref{laxcurlylax} and
\eqref{laxbarcurlylax}. 

This is the setup we need in order to adapt the general construction
method of section~\ref{genconstr} to the superalgebra case.
Importantly, the classification of the invariants using permutations
and the decomposition of these permutations into transpositions does
not make any reference to the specific symmetry algebra, let it be
$\mathfrak{gl}(2)$ or $\mathfrak{gl}(N|M)$. The form of the vacuum
$|\mathbf{0}\rangle$ in \eqref{multifock}, the operators
$\mathcal{B}_{i j}(u)$ specified in \eqref{operatorB}, as well as the
key relation \eqref{YBElike} remain unchanged. Furthermore, all
formulas in section~\ref{genconstr} already contain $N-M$ at the
appropriate places, and already pertain to the $\mathfrak{gl}(N|M)$
case. Hence the construction of Yangian invariants carries over to the
case of compact $\mathfrak{gl}(N|M)$ representations, where the
generators \eqref{supergensym} and \eqref{supergendual} are given in
terms of a single family of superoscillators.

For the moment we stay with these compact representations, and discuss
an integral realization of the operators $\mathcal{B}_{i j}(u)$. Soon
this realization will be the basis for a formal transition to scattering amplitudes. As was observed in
\cite{Chicherin:2013ora} (see also \cite{Frassek:2013xza}), one may formally rewrite the intertwiner $\mathcal{B}_{i j}(u)$ in \eqref{operatorB} for arbitrary complex numbers\footnote{For $u=-1,-2,
  \ldots$ one needs to take a limit, as the Gamma function diverges
  and the integral tends to zero since the integrand becomes
  completely analytic inside $\mathcal{C}$.} $u \in \mathbb{C}$ as
\begin{equation}
  \label{integraloperatorB}
  \mathcal{B}_{i j}(u)=
  (-\oscwb^j \cdot \oscw^i)^u 
  =-\frac{\Gamma(u+1)}{2 \pi i}\int_{\mathcal{C}}\frac{d\alpha}{(-\alpha)^{1+u}}
  e^{\alpha\, \oscwb^j \cdot \oscw^i}\,,
\end{equation}
where the Hankel contour $\mathcal{C}$ goes counterclockwise around
the cut (for $u\notin \mathbb{Z}$) of the function $(-\alpha)^{1+u}$
defined to lie between its branch points at 0 and $\infty$.
Now we make a further notational change, and realize the $\oscw$ oscillators in terms of ``supertwistor'' variables $\mathcal{W}$
\begin{align}
  \oscwb_a\leftrightarrow\mathcal{W}_a\,,\qquad
  \oscw_a\leftrightarrow\partial_{\mathcal{W}_a}\,,\quad
  |0\rangle\leftrightarrow 1\,,\quad
  |\bar{0}\rangle\leftrightarrow\delta^{N|M}({\mathcal{W}})\,.
\end{align}
In this realization the vacuum state \eqref{multifock} in the construction of invariants becomes
\begin{align}\label{multidelta}
  |\mathbf{0}\rangle=
  \prod_{i=1}^k\delta^{N|M}(\mathcal{W}^i)\,.
\end{align}
The operators $\mathcal{B}_{i j}(u)$ in \eqref{integraloperatorB} then read 
\begin{equation}
  \label{integraloperatorBamp}
  \mathcal{B}_{i j}(u)=
  -\frac{\Gamma(u+1)}{2 \pi i}\int_{\mathcal{C}}\frac{d\alpha}{(-\alpha)^{1+u}}
  e^{\alpha\, \mathcal{W}^j \cdot \partial_{\mathcal{W}^i}}\,.
\end{equation}

Note that for $u=s\in \mathbb{N}_0$ the cut in the complex
$\alpha$-plane disappears, and the only singularity inside
$\mathcal{C}$ is a pole at $\alpha=0$. In this case, the Hankel
contour may be collapsed into a counterclockwise circle around the
pole, and \eqref{integraloperatorBamp} simplifies to
\begin{equation}\label{hankelcollapsed}
\mathcal{B}_{i j}(s)
=\frac{(-1)^s\,s!}{2 \pi i}\oint\frac{d\alpha}{\alpha^{1+s}}
e^{\alpha\, \mathcal{W}^j \cdot \partial_{\mathcal{W}^i}}
=(-\mathcal{W}^j \cdot \partial_{\mathcal{W}^i})^s\,,
\end{equation}
which is essentially the expression, here in terms of twistor variables, given in (5.11) of \cite{Frassek:2013xza} for compact representations. We may then express, as in (5.21) of \cite{Frassek:2013xza}, e.g.\ \eqref{psi21} in twistor variables
\begin{equation}\label{psi21continued}
|\Psi\rangle_{2,1}
\propto \oint \frac{d\alpha}{\alpha^{1+s_2}}
\delta^{2|0}(\mathcal{W}^1+\alpha\, \mathcal{W}^2)\,,
\end{equation}
where we used the representation \eqref{multidelta} of the vacuum. However, for more complicated compact invariants such as \eqref{42fromB}, where the intertwiners depend on complex parameters, we should use the integral representations \eqref{integraloperatorBamp} employing the Hankel contours.

Now, as already pointed out in \cite{Frassek:2013xza}, with a small further modification the compact formalism is easily modified to also apply to the non-compact $\mathfrak{gl}(4|4)$ Yangian invariants appearing in the $\mathcal{N}=4$ tree-level scattering problem, which are expressed in terms of formal Gra{\ss}mannian contour integrals \cite{ArkaniHamed:2012nw}. As the procedure for building invariants is entirely algebraic, and reality conditions, conjugation properties of the operators, and the norms of the states are never considered, the construction immediately carries over to the amplitude problem. The price one pays is merely that the contour integrations and the delta functions, which are well-defined in the compact case (cf.\ appendix A of  \cite{Frassek:2013xza}) become somewhat formal in the scattering problem, see again \cite{ArkaniHamed:2012nw}. Let us now illustrate the method in some examples.

\section{Sample Invariants}\label{Sec:Examples}
We will illustrate our construction in the first few 
cases. As we elaborated before, all the invariants are labelled by
permutations. Here we restrict our discussion to the permutations $\sigma_{n,k}(i)=i+k \mbox{ mod } n$ that are relevant
for the top cells of the positive Gra{\ss}mannian $G(n,k)$. For the corresponding invariants we use the shorthand notation
$|\Psi\rangle_{n,k}=|\Psi\rangle_{\sigma_{n,k}}$. Each invariant
may be rewritten into integral form using \eqref{integraloperatorBamp}. However, we will drop the reference to the Hankel contours, in order to stay general. There are
at least three natural sets of variables in which the integral
representation can be expressed. Directly using \eqref{integraloperatorBamp}, we may
write it employing variables $\alpha_p$, which are related to the
decomposition into BCFW bridges \eqref{decomposition}, see also
\cite{ArkaniHamed:2012nw}. Another set of variables is given by
the entries of the matrix $C=(c_{ij})^{i=1,\ldots,k}_{j=1,\ldots,n}$ making
its appearance in the Gra\ss mannian integral formulation studied in
\cite{ArkaniHamed:2009dn}. The last form is given in terms of the face
variables $f_i$ introduced in \cite{Postnikov:2006kva}.

As was already mentioned in the previous section, the ensuing formulas are valid for both the Yangian invariants of compact $\mathfrak{gl}(N|M)$ algebras as well as for the amplitude problem of $\mathcal{N}=4$ SYM with $\mathfrak{gl}(4|4)$ symmetry.

\subsection{n=2, k=1}
There is only one non-trivial two-point invariant, for which $k=1$. The permutation reduces just to a single transposition
\begin{equation}
\sigma_{2,1}=\left( \begin{tabular}{ccc}1&2\\2&1\end{tabular}\right)=(12)\,,
\end{equation}
The invariant \eqref{ansatzyi} in that case is given by
\begin{align}
  |\Psi\rangle_{2,1}=\mathcal{B}_{12}(y_1-y_2)|\mathbf{0}\rangle=
  \mathcal{B}_{12}(\mathfrak{s}_2)|\mathbf{0}\rangle\propto
  \int \frac{d\alpha_1}{\alpha_1^{1+\mathfrak{s}_2}}\delta^{N|M}(\mathcal{W}^1+\alpha_1 \mathcal{W}^2)\,,
\end{align}
where we used the fact that $y_1=y_2+\mathfrak{s}_2$.

\begin{figure}[ht]
\begin{center}
\scalebox{0.35}{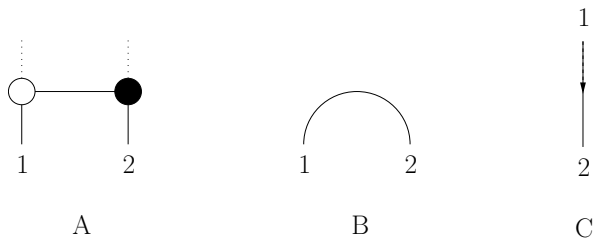}
\end{center}
\caption{Two-point Yangian invariant. A) Transposition decomposition; B) wilted on-shell diagram; C) on-shell diagram with a perfect orientation.}
\label{Fig:n2k1}
\end{figure}

There are at least three distinguished graphical representations for
each invariant. The first one, as proposed in
\cite{Chicherin:2013ora}, comes from the identification of
$\mathcal{B}_{ij}$ with a BCFW bridge. Each such bridge can be
depicted as a composition of one white and one black three-point
vertex. Then the invariant can be drawn\footnote{We thank Yvonne Geyer
  for bringing this construction to our attention in the case of
  undeformed amplitudes.} as in Figure~\ref{Fig:n2k1}A. When we remove
all dotted lines and replace all vertices with only two solid lines by
a single solid line, we end up with a graphical representation
analogous to the one in \cite{Frassek:2013xza}, as depicted in
Figure~\ref{Fig:n2k1}B. It is now easy to translate the latter into an
on-shell diagram as in Figure~\ref{Fig:n2k1}C, which obviously is
rather trivial in the case of the two-point invariant.

\subsection{n=3, k=1}
\begin{figure}[ht]
\begin{center}
\scalebox{0.35}{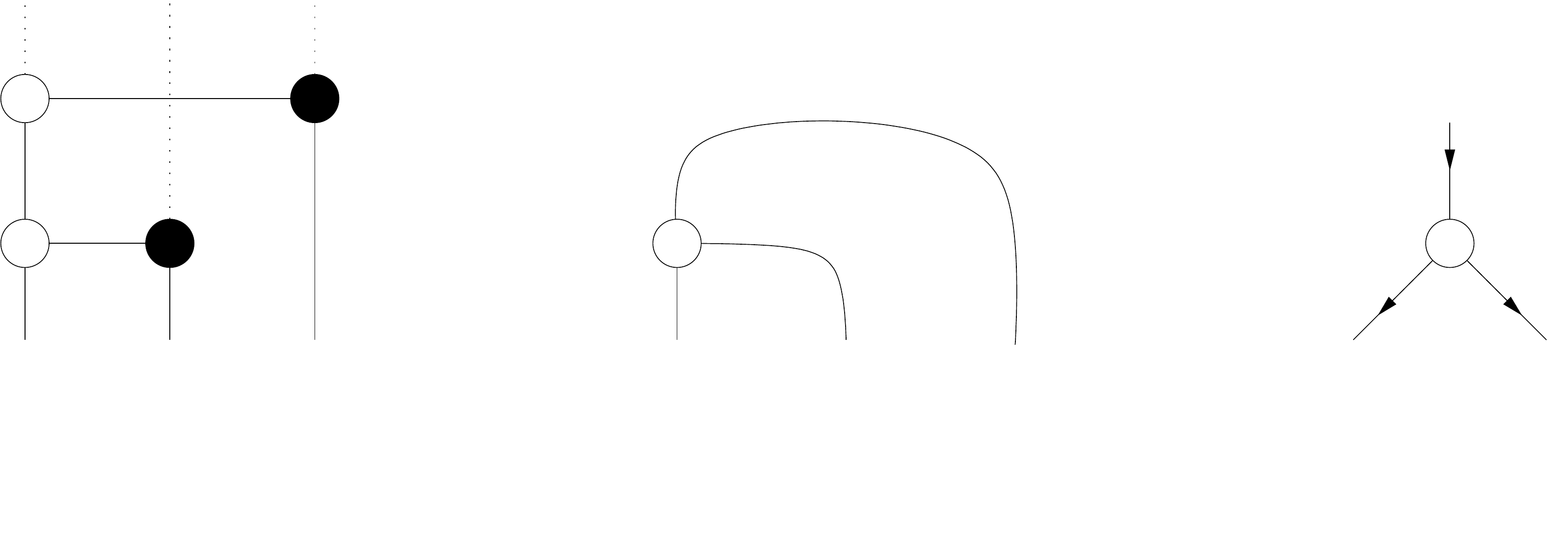}
\end{center}
\caption{Three-point $\overline{\mbox{MHV}}$ Yangian invariant. A) Transposition decomposition; B) wilted on-shell diagram; C) on-shell diagram with a perfect orientation.}
\label{Fig:n3k1}
\end{figure}

For the three-particle invariant there are two non-trivial values of $k$. Let us first take the permutation
\begin{equation}
\sigma_{3,1}=\left( \begin{tabular}{ccc}1&2&3\\2&3&1\end{tabular}\right)=(13)(12)\,,
\end{equation}
for which $k=1$. We specified here also the decomposition of this
permutation into transpositions. The invariant \eqref{ansatzyi} is
given by
\begin{align}
|\Psi\rangle_{3,1}&=
\mathcal{B}_{12}(y_1-y_2)\mathcal{B}_{13}(y_2-y_3)|\mathbf{0}\rangle=\mathcal{B}_{12}(\mathfrak{s}_2)\mathcal{B}_{13}(\mathfrak{s}_3)|\mathbf{0}\rangle\\ 
&\propto\int \frac{d\alpha_{1} d\alpha_{2}}{\alpha_{1}^{1+\mathfrak{s}_2}\alpha_{2}^{1+\mathfrak{s}_3}}
\delta^{N|M}(\mathcal{W}^1+\alpha_{1} \mathcal{W}^2+\alpha_{2} \mathcal{W}^3)\\\label{invariant31}
&\propto\int \frac{dc_{12} dc_{13}}{c_{12}^{1+\mathfrak{s}_2}c_{13}^{1+\mathfrak{s}_3}}\delta^{N|M}(\mathcal{W}^1+c_{12} \mathcal{W}^2+c_{13} \mathcal{W}^3)\,.
\end{align}
This is exactly the deformed three-point $\overline{\mbox{MHV}}$ amplitude introduced in \cite{Ferro:2012xw}. See Figure~\ref{Fig:n3k1} for the graphical representation of $|\Psi\rangle_{3,1}$.

\subsection{n=3, k=2}
\begin{figure}[ht]
\begin{center}
\scalebox{0.35}{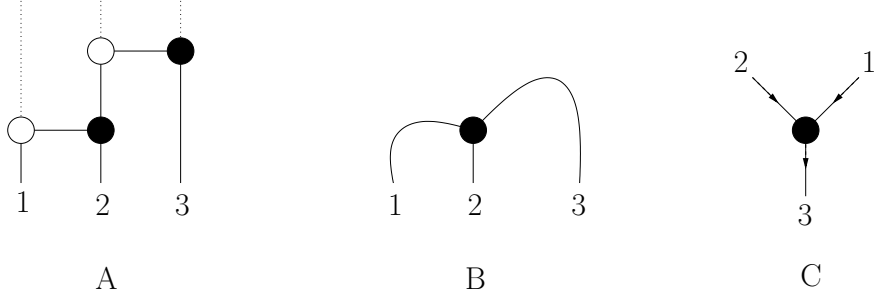}
\end{center}
\caption{Three-point MHV Yangian invariant. A) Transposition decomposition; B) wilted on-shell diagram; C) on-shell diagram with a perfect orientation.}
\label{Fig:n3k2}
\end{figure}

For the case of the three-particle invariant with $k=2$ we have the permutation 
\begin{equation}
\sigma_{3,2}=\left( \begin{tabular}{ccc}1&2&3\\3&1&2\end{tabular}\right)=(23)(12)\,.
\end{equation}
The invariant is given by
\begin{align}
|\Psi\rangle_{3,2}&=\mathcal{B}_{12}(y_{1}-y_{2})\mathcal{B}_{23}(y_{1}-y_3)|\mathbf{0}\rangle
=\mathcal{B}_{12}(-\mathfrak{s}_1)\mathcal{B}_{23}(\mathfrak{s}_3)|\mathbf{0}\rangle\\&\propto
\int \frac{d\alpha_1 d\alpha_2}{\alpha_1^{1-\mathfrak{s}_1}\alpha_2^{1+\mathfrak{s}_3}}\delta^{N|M}(\mathcal{W}^1+
\alpha_1 \mathcal{W}^2)\delta^{N|M}(\mathcal{W}^2+\alpha_2 \mathcal{W}^3)\\
&\propto\int \frac{d c_{13} d c_{23}}{ c_{13}^{1-\mathfrak{s}_1} c_{23}^{1-\mathfrak{s}_2}}\delta^{N|M}(\mathcal{W}^1+ c_{13} \mathcal{W}^3)\delta^{N|M}(\mathcal{W}^2+ c_{23} \mathcal{W}^3)\,.
\end{align}
This is again the deformed three-point MHV amplitude found in \cite{Ferro:2012xw}.  Together with \eqref{invariant31} it is the building block for all deformations of on-shell diagrams, and subsequently all deformed tree-level amplitudes. For the graphical representation see Figure~\ref{Fig:n3k2}.

\subsection{n=4, k=2}
The four-point invariant with $k=2$ is the first one which, interestingly, cannot be written solely by using representation labels. It corresponds to the deformation of the four-point tree amplitude obtained in \cite{Ferro:2013dga} and depends on a spectral parameter $z$. Let us show how it arises in the context of this paper. The relevant permutation is
\begin{equation}
\sigma_{4,2}=\left( \begin{tabular}{cccc}1&2&3&4\\3&4&1&2\end{tabular}\right)=(24)(12)(23)(12)\,.
\end{equation}
and the invariant is given by
\begin{align}
|\Psi\rangle_{4,2}&=\mathcal{B}_{12}(y_{1}-y_2)\mathcal{B}_{23}(y_1-y_3)\mathcal{B}_{12}(y_2-y_3)\mathcal{B}_{24}(y_2-y_4)|\mathbf{0}\rangle\\
&=\mathcal{B}_{12}(z)\mathcal{B}_{23}(-\mathfrak{s}_1)\mathcal{B}_{12}(-z-\mathfrak{s}_1)\mathcal{B}_{24}(-\mathfrak{s}_2)|\mathbf{0}\rangle\label{42-invariantb}\\
\label{42-invariant}
&\propto\int \frac{df_1 df_2 df_3 df_4}{f_1^{1-\mathfrak{s}_1}f_2^{1-\mathfrak{s}_1-\mathfrak{s}_2}f_3^{1-z-\mathfrak{s}_1}f_{4}^{1-\mathfrak{s}_2}}
\,\delta^{N|M}(\mathcal{W}^1+f_1 f_2 \mathcal{W}^3+(1+f_3)f_1 f_2 f_4 \mathcal{W}^4)\\
&\hspace{1cm}\delta^{N|M}(\mathcal{W}^2+f_2 \mathcal{W}^3+f_2 f_4 \mathcal{W}^4)\,,
\end{align}
where we defined $z=y_{1}-y_{2}$. The form \eqref{42-invariantb} of this invariant was already mentioned in \eqref{42fromB}. Its integral representation \eqref{42-invariant} exactly reproduces the four-point deformed amplitude in the form derived in \cite{Ferro:2012xw}, see also \cite{Ferro:2013dga}. Note that a somewhat different looking form involving a hypergeometric function in the integrand of the harmonic R-matrix was given in \cite{Frassek:2013xza}.

\begin{figure}[ht]
\begin{center}
\scalebox{0.28}{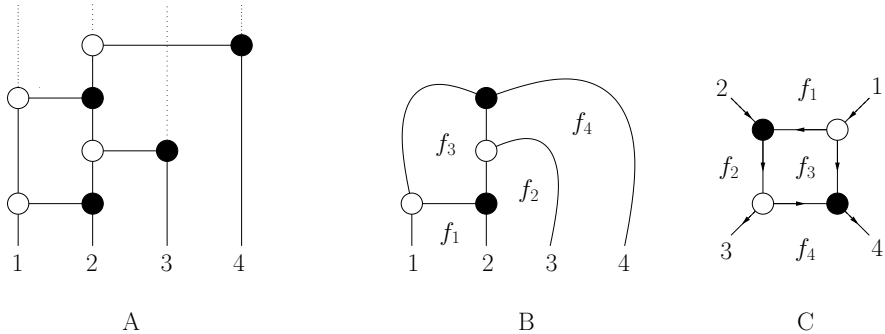}
\end{center}
\caption{Four-point MHV Yangian invariant. A) Transposition decomposition; B) wilted on-shell diagram; C) on-shell diagram with a perfect orientation.}
\end{figure}

\subsection{n=5, k=2}
In the five particle case the permutation
\begin{equation}
\sigma_{5,2}=\left( \begin{tabular}{ccccc}1&2&3&4&5\\3&4&5&1&2\end{tabular}\right)=(25)(12)(24)(12)(23)(12)\,
\end{equation}
with $k=2$ leads to the invariant
\begin{align}
|\Psi\rangle_{5,2}&=\mathcal{B}_{12}(y_{12})\mathcal{B}_{23}(y_{13})\mathcal{B}_{12}(y_{23})\mathcal{B}_{24}(y_{24})\mathcal{B}_{12}(y_{34})\mathcal{B}_{25}(y_{35})|\mathbf{0}\rangle\\
&=\mathcal{B}_{12}(-\mathfrak{s}_1-\mathfrak{s}_4)\mathcal{B}_{23}(\mathfrak{s}_3)\mathcal{B}_{12}(-\mathfrak{s}_2-\mathfrak{s}_5)\mathcal{B}_{24}(\mathfrak{s}_4)\mathcal{B}_{12}(-\mathfrak{s}_1-\mathfrak{s}_3)\mathcal{B}_{25}(\mathfrak{s}_5)|\mathbf{0}\rangle\\
&\propto\int \frac{df_1 df_2 df_3 df_4 df_5 df_6}{
f_1^{1-\mathfrak{s}_1}f_2^{1-\mathfrak{s}_1-\mathfrak{s}_2}f_{3}^{1+\mathfrak{s}_4}f_4^{1+\mathfrak{s}_4+\mathfrak{s}_5}f_5^{1-\mathfrak{s}_1-\mathfrak{s}_3}f_6^{1+\mathfrak{s}_5}}
\\&\qquad\delta^{N|M}(\mathcal{W}^1+f_1 f_2\mathcal{W}^3+(1+f_3)f_1 f_2 f_4 \mathcal{W}^4+(1+f_3+f_3 f_5)f_1 f_2 f_4 f_6\mathcal{W}^5)\\
&\qquad\delta^{N|M}(\mathcal{W}^2+f_2 \mathcal{W}^3+f_2 f_4 \mathcal{W}^4+f_2 f_4 f_6 \mathcal{W}^5)\,,
\end{align}
where we abbreviated $y_{ij}=y_i-y_j$. This provides a deformation of the five-point MHV amplitude. Notice that it is fully determined just using representation labels.
\begin{figure}[ht]
\begin{center}
\scalebox{0.28}{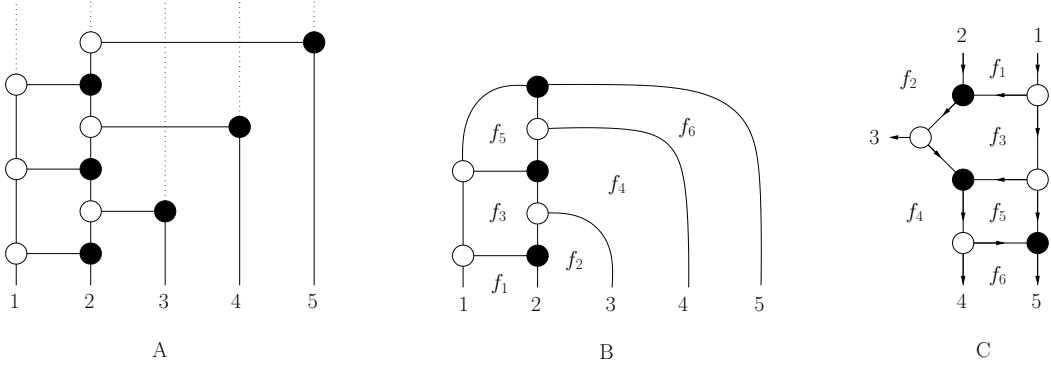}
\end{center}
\caption{Five-point MHV Yangian invariant. A) Transposition decomposition; B) wilted on-shell diagram; C) on-shell diagram with a perfect orientation.}
\end{figure}

\subsection{n=5, k=3}
A five particle permutation with $k=3$ is given by
\begin{equation}
\sigma_{5,3}=\left( \begin{tabular}{ccccc}1&2&3&4&5\\4&5&1&2&3\end{tabular}\right)=(35)(23)(34)(12)(23)(12)\,.
\end{equation}
This yields the invariant
\begin{align}
|\Psi\rangle_{5,3}&=\mathcal{B}_{12}(y_{12})\mathcal{B}_{23}(y_{13})\mathcal{B}_{12}(y_{23})\mathcal{B}_{34}(y_{14})\mathcal{B}_{23}(y_{24})\mathcal{B}_{35}(y_{25})|\mathbf{0}\rangle\\
&=\mathcal{B}_{12}(\mathfrak{s}_2+\mathfrak{s}_4)\mathcal{B}_{23}(-\mathfrak{s}_1)\mathcal{B}_{12}(\mathfrak{s}_3+\mathfrak{s}_5)\mathcal{B}_{34}(\mathfrak{s}_4)\mathcal{B}_{23}(-\mathfrak{s}_2)\mathcal{B}_{35}(\mathfrak{s}_5)|\mathbf{0}\rangle\\
&\propto\int \frac{df_1 df_2 df_3 df_4 df_5 df_6}{
 f_1^{1-\mathfrak{s}_1}f_2^{1-\mathfrak{s}_1-\mathfrak{s}_2}f_{3}^{1+\mathfrak{s}_3+\mathfrak{s}_5}f_4^{1+\mathfrak{s}_4+\mathfrak{s}_5}f_5^{1-\mathfrak{s}_2}f_6^{1+\mathfrak{s}_5}}\\
&\qquad\delta^{N|M}(\mathcal{W}^1+f_1 f_2 f_4 \mathcal{W}^4+f_1 f_2 f_4 f_6(1+f_5+f_3 f_5)\mathcal{W}^5)\\
&\qquad\delta^{N|M}(\mathcal{W}^2+f_2 f_4\mathcal{W}^4+f_2 f_4 f_6(1+f_5)\mathcal{W}^5)\\
&\qquad\delta^{N|M}(\mathcal{W}^3+f_4\mathcal{W}^4+f_4 f_6 \mathcal{W}^5)\,,
\end{align}
which is a deformation of the five-point $\overline{\mbox{MHV}}$ amplitude.

\begin{figure}[ht]
\begin{center}
\scalebox{0.28}{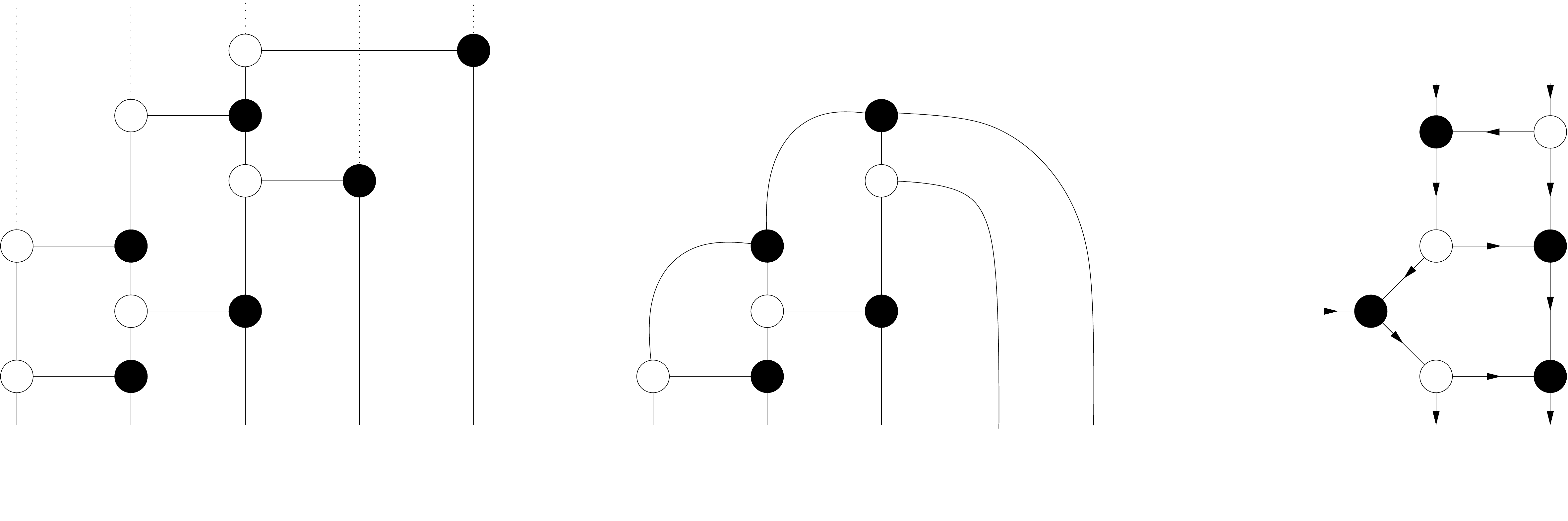}
\end{center}
\caption{Five-point $\overline{\mbox{MHV}}$ Yangian invariant. A) Transposition decomposition; B) wilted on-shell diagram; C) on-shell diagram with a perfect orientation.}
\end{figure}

\section{Summary and Outlook}
In this paper we provided a full classification of Yangian invariants relevant to tree-level scattering amplitudes in $\mathcal{N}=4$ SYM. Our method combines the idea of deformation of on-shell diagrams \cite{Ferro:2013dga} with the QISM as proposed in \cite{Frassek:2013xza} and \cite{Chicherin:2013ora}. It gives a constructive way to build such invariants and provides a link to powerful Bethe Ansatz methods. It also introduces natural variables $v^{\pm}$, as studied in details by \cite{Beisert:2014qba}, which are very reminiscent of the Zhukovsky variables $x^\pm$ that play an important role in the all-loop solution of the spectral problem \cite{Beisert:2005fw}. This leads to the hope that these variables will provide a good starting point for suitable generalizations of Yangian invariants to scattering amplitudes at higher loops.  


\section*{Acknowledgments}
We thank Livia Ferro, Rouven Frassek, Yvonne Geyer, Yumi Ko and Arthur
Lipstein for useful discussions. This research is supported in part
by the SFB 647 \emph{``Raum-Zeit-Materie. Analytische und Geometrische
  Strukturen''} and the Marie Curie network GATIS
(\texttt{\href{http://gatis.desy.eu}{gatis.desy.eu}}) of the European
Union’s Seventh Framework Programme FP7/2007-2013/ under REA Grant
Agreement No 317089. N.K.\ is supported by a
\emph{Promotionsstipendium} of the \emph{Studienstiftung des Deutschen
  Volkes}, and receives partial support by the GK 1504 \emph{``Masse,
  Spektrum, Symmetrie''}.


\appendix

\section{Other Form of Intertwining Relation}
\label{interwining-relation}

In this appendix we work out the precise connection between the
commutation relation \eqref{YBElike}, which is the primary tool to
construct Yangian invariants in this paper, and the corresponding
relation in \cite{Chicherin:2013ora}.

We start by multiplying \eqref{YBElike} from the left with
$\mathcal{L}_j(-u,-y_j-1)\mathcal{L}_i(-u,-y_i-1)$ and from the right
with $\mathcal{L}_j(-u,-y_i-1)\mathcal{L}_i(-u,-y_j-1)$. In the
resulting equation products of Lax operators acting on the same space
are eliminated using the ``unitarity condition''
\begin{align}
  \label{eq:laxunitarity}
  \mathcal{L}_i(u,v)\mathcal{L}_i(-u,-v-1+\mathfrak{C}_i)
  =(u-v)(-u+v+1-\mathfrak{C}_i)+\mathfrak{C}_i\,.
\end{align}
Then the equation is simplified further employing
\begin{align}
  \label{eq:commrelbc}
  \mathfrak{C}_i\mathcal{B}_{ij}(u)=\mathcal{B}_{ij}(u)(\mathfrak{C}_i-u)\,,\qquad
  \mathfrak{C}_j\mathcal{B}_{ij}(u)=\mathcal{B}_{ij}(u)(\mathfrak{C}_j+u)\,.
\end{align}
Finally, relabeling $u\mapsto -u$, $y_i\mapsto -y_j-1$, $y_j\mapsto
-y_i-1$ leads to the commutation relation used in
\cite{Chicherin:2013ora}, which does not contain the operators
$\mathfrak{C}_i$ and $\mathfrak{C}_j$,
\begin{align}
  \label{eq:fundrel-chicherin}
  \mathcal{B}_{ij}(y_i-y_j)\mathcal{L}_j(u,y_j)\mathcal{L}_i(u,y_i)
  =
  \mathcal{L}_j(u,y_i)\mathcal{L}_i(u,y_j)\mathcal{B}_{ij}(y_i-y_j)\,.
\end{align}


\bibliographystyle{nb}
\bibliography{bibliography}

\begin{thebibliography}{10}
\ifx\href\asklfhas\newcommand{\href}[2]{#2}\fi
\ifx\arxivref\asklfhas\newcommand{\arxivref}[2]{\href{http://arxiv.org/abs/#1}%
{#2}}\fi
\ifx\doiref\asklfhas\newcommand{\doiref}[2]{\href{http://dx.doi.org/#1}{#2}}\fi
\raggedright
\small
\parskip 0pt

\bibitem{Beisert:2010jr}
N.~Beisert, C.~Ahn, L.~F.~Alday, Z.~Bajnok, J.~M.~Drummond et~al.,
\textit{``{Review of AdS/CFT Integrability: An Overview}''},
\textsf{\doiref{10.1007/s11005-011-0529-2}{Lett.~Math.~Phys.~99,~3~(2012)}},
\texttt{\arxivref{1012.3982}{arxiv:1012.3982}}.

\bibitem{Gromov:2013pga}
N.~Gromov, V.~Kazakov, S.~Leurent and D.~Volin,
\textit{``{Quantum Spectral Curve for $AdS_5/CFT_4$}''},
\textsf{\doiref{10.1103/PhysRevLett.112.011602}{Phys.~Rev.~Lett.~112,~011602~(%
2014)}},
\texttt{\arxivref{1305.1939}{arxiv:1305.1939}}.

\bibitem{Alday:2010vh}
L.~F.~Alday, J.~Maldacena, A.~Sever and P.~Vieira,
\textit{``{Y-System for Scattering Amplitudes}''},
\textsf{\doiref{10.1088/1751-8113/43/48/485401}{J.Phys.~A43,~485401~(2010)}},
\texttt{\arxivref{1002.2459}{arxiv:1002.2459}}.

\bibitem{Basso:2013vsa}
B.~Basso, A.~Sever and P.~Vieira,
\textit{``{Spacetime and Flux Tube S-Matrices at Finite Coupling for
  $\mathcal{N}=4$ Supersymmetric Yang-Mills Theory}''},
\textsf{\doiref{10.1103/PhysRevLett.111.091602}{Phys.~Rev.~Lett.~111,~091602~(%
2013)}},
\texttt{\arxivref{1303.1396}{arxiv:1303.1396}}.

\bibitem{Basso:2013aha}
B.~Basso, A.~Sever and P.~Vieira,
\textit{``{Space-time S-Matrix and Flux Tube S-Matrix II. Extracting and
  Matching Data}''},
\textsf{\doiref{10.1007/JHEP01(2014)008}{JHEP~1401,~008~(2014)}},
\texttt{\arxivref{1306.2058}{arxiv:1306.2058}}.

\bibitem{Basso:2014koa}
B.~Basso, A.~Sever and P.~Vieira,
\textit{``{Space-time S-Matrix and Flux-tube S-Matrix III. The Two-particle
  Contributions}''},
\texttt{\arxivref{1402.3307}{arxiv:1402.3307}}.

\bibitem{Drummond:2009fd}
J.~M.~Drummond, J.~M.~Henn and J.~Plefka,
\textit{``{Yangian Symmetry of Scattering Amplitudes in
  {$\mathcal{N}=\mathord{}$4} Super Yang-Mills theory}''},
\textsf{\doiref{10.1088/1126-6708/2009/05/046}{JHEP~0905,~046~(2009)}},
\texttt{\arxivref{0902.2987}{arxiv:0902.2987}}.

\bibitem{Drummond:2008vq}
J.~M.~Drummond, J.~Henn, G.~P.~Korchemsky and E.~Sokatchev,
\textit{``{Dual Superconformal Symmetry of Scattering Amplitudes in
  {$\mathcal{N}=4$} Super-Yang-Mills Theory}''},
\textsf{\doiref{10.1016/j.nuclphysb.2009.11.022}{Nucl.~Phys.~B828,~317~(2010)}%
},
\texttt{\arxivref{0807.1095}{arxiv:0807.1095}}.

\bibitem{Ferro:2012xw}
L.~Ferro, T.~Lukowski, C.~Meneghelli, J.~Plefka and M.~Staudacher,
\textit{``{Harmonic R-Matrices for Scattering Amplitudes and Spectral
  Regularization}''},
\textsf{\doiref{10.1103/PhysRevLett.110.121602}{Phys.~Rev.~Lett.~110,~121602~(%
2013)}},
\texttt{\arxivref{1212.0850}{arxiv:1212.0850}}.

\bibitem{Ferro:2013dga}
L.~Ferro, T.~Łukowski, C.~Meneghelli, J.~Plefka and M.~Staudacher,
\textit{``{Spectral Parameters for Scattering Amplitudes in $\mathcal{N}=4$
  Super Yang-Mills Theory}''},
\textsf{\doiref{10.1007/JHEP01(2014)094}{JHEP~1401,~094~(2014)}},
\texttt{\arxivref{1308.3494}{arxiv:1308.3494}}.

\bibitem{ArkaniHamed:2012nw}
N.~Arkani-Hamed, J.~L.~Bourjaily, F.~Cachazo, A.~B.~Goncharov, A.~Postnikov
  et~al.,
\textit{``{Scattering Amplitudes and the Positive Grassmannian}''},
\texttt{\arxivref{1212.5605}{arxiv:1212.5605}}.

\bibitem{Britto:2005fq}
R.~Britto, F.~Cachazo, B.~Feng and E.~Witten,
\textit{``Direct Proof of Tree-level Recursion Relation in Yang-Mills
  Theory''},
\textsf{\doiref{10.1103/PhysRevLett.94.181602}{Phys.~Rev.~Lett.~94,~181602~(20%
05)}},
\texttt{\arxivref{hep-th/0501052}{hep-th/0501052}}.

\bibitem{Postnikov:2006kva}
A.~Postnikov,
\textit{``{Total Positivity, Grassmannians, and Networks}''},
\texttt{\arxivref{math/0609764}{math/0609764}}.

\bibitem{Beisert:2014qba}
N.~Beisert, J.~Broedel and M.~Rosso,
\textit{``{On Yangian-invariant Regularisation of Deformed On-shell Diagrams in
  $\mathcal{N}=4$ Super-Yang-Mills Theory}''},
\texttt{\arxivref{1401.7274}{arxiv:1401.7274}}.

\bibitem{Frassek:2013xza}
R.~Frassek, N.~Kanning, Y.~Ko and M.~Staudacher,
\textit{``{Bethe Ansatz for Yangian Invariants: Towards Super Yang-Mills
  Scattering Amplitudes}''},
\texttt{\arxivref{1312.1693}{arxiv:1312.1693}}.

\bibitem{Chicherin:2013ora}
D.~Chicherin, S.~Derkachov and R.~Kirschner,
\textit{``{Yang-Baxter Operators and Scattering Amplitudes in $\mathcal{N}=4$
  Super-Yang-Mills Theory}''},
\textsf{\doiref{10.1016/j.nuclphysb.2014.02.016}{Nucl.~Phys.~B881,~467~(2014)}%
},
\texttt{\arxivref{1309.5748}{arxiv:1309.5748}}.

\bibitem{Broedel:2014pia}
J.~Broedel, M.~de~Leeuw and M.~Rosso,
\textit{``{A Dictionary Between R-Operators, On-shell Graphs and Yangian
  Algebras}''},
\texttt{\arxivref{1403.3670}{arxiv:1403.3670}}.

\bibitem{Bourjaily:2012gy}
J.~L.~Bourjaily,
\textit{``{Positroids, Plabic Graphs, and Scattering Amplitudes in
  Mathematica}''},
\texttt{\arxivref{1212.6974}{arxiv:1212.6974}}.

\bibitem{Kirschner:2013ila}
R.~Kirschner,
\textit{``{Integrable Chains with Jordan-Schwinger Representations}''},
\textsf{\doiref{10.1088/1742-6596/411/1/012018}{J.~Phys.~Conf.~Ser.~411,~01201%
8~(2013)}}.

\bibitem{ArkaniHamed:2009dn}
N.~Arkani-Hamed, F.~Cachazo, C.~Cheung and J.~Kaplan,
\textit{``{A Duality For The S Matrix}''},
\textsf{\doiref{10.1007/JHEP03(2010)020}{JHEP~1003,~020~(2010)}},
\texttt{\arxivref{0907.5418}{arxiv:0907.5418}}.

\bibitem{Beisert:2005fw}
N.~Beisert and M.~Staudacher,
\textit{``Long-Range $\mathfrak{psu}(2,2|4)$ Bethe Ansaetze for Gauge Theory
  and Strings''},
\textsf{\doiref{10.1016/j.nuclphysb.2005.06.038}{Nucl.~Phys.~B727,~1~(2005)}},
\texttt{\arxivref{hep-th/0504190}{hep-th/0504190}}.

\end{thebibliography}
\end{document}